\documentclass{article}
\usepackage{amsfonts}
\usepackage{amssymb}
\usepackage{amsmath}
\usepackage{amsthm}
\usepackage{mathtools}

\begin{document}

\title{More on Bianchi I spacetimes and $f(T)$ gravity}

\author{Alexey Golovnev${}^{1}$, Mustafa M. Hemida${}^{2}$\\
{\small ${}^{1}${\it Centre for Theoretical Physics, the British University in Egypt,}}\\ 
{\small {\it BUE 11837, El Sherouk City, Cairo Governorate, Egypt}}\\
{\small agolovnev@yandex.ru}\\
{\small ${}^{2}${\it Physics of Universe Program, Zewail City of Science, Technology and Innovation,}}\\ 
{\small {\it October Gardens, Giza 12578, Egypt
}}\\
{\small s-mustafa.elsayed@zewailcity.edu.eg}
}
\date{}

\maketitle

\begin{abstract}

Bianchi I cosmological solutions in $f(T)$ gravity are discussed. We start from diagonal metrics and tetrads and show that their dynamical equations are pretty much tractable analytically, with a possible arena for physical applications. Then we derive a very bad unpredictability of the teleparallel connection in these configurations. Namely, even for the simple isotropic Friedmann universes, one might apply an arbitrary time-dependent spatial rotation to the standard tetrad 
without changing anything in the cosmological equations.
 
\end{abstract}

\section{Introduction}

Nowadays, various modified gravity models are a popular field of research, with lots of motivation for them including the cosmological tensions. The modifications can be different and range from very little additions to the classical general relativity (GR) to some profound changes in the foundations. In particular, teleparallel approaches do change the very geometry in hand. On top of the metric geometry, there is also yet another connection in the spacetime manifold, a connection that is flat. It might be in terms of either torsion or nonmetricity, or both. 

In this paper, we deal with the more classical, torsion-based approach dating back to Einstein himself. One of the most popular classes of modified theories in the metric teleparallel framework, i.e. torsion-based with no nonmetricity, is $f(T)$ gravity \cite{fT}. Note that we denote its torsion scalar $T$ as $\mathbb T$. These theories are being widely used for cosmological model building, even though there are many problems of them on the foundational side \cite{meiss}. We shall not delve in discussions on these very interesting topics in the current work.

Neglecting the problems for a moment, we would like to show that the Bianchi type I cosmological solutions are easy to construct in $f(\mathbb T)$ gravity, and in many respects they can be investigated analytically, unlike many accounts one might see in the available literature. We will give some details on how it can be done, from both Lagrangian and Hamiltonian viewpoints, and discuss a possible application. In the end of the paper, we will come back to the foundational problems again and see that we are too much free in choosing the tetrad for the Bianchi I solutions, with the freedom implying an unpredictable evolution of the teleparallel connection.

We consider the $f(\mathbb T)$ theory in the pure tetrad approach. Equivalently, one can say that we have fixed the Weitzenb{\"o}ck gauge (vanishing spin connection) in the covariant approah. Geometrically, it means that we describe the teleparallel structure in terms of an objectively preferred tetrad \cite{megeom}. It is understood as a tangent space basis composed of covariantly constant vectors, $\bigtriangledown^{(\mathrm{TP})}_{\mu} e^{\nu}_A = \partial_{\mu}e^{\nu}_A + \Gamma^{\nu}_{\mu\alpha} e^{\alpha}_A =0$, in terms of the teleparallel (TP) connection $\Gamma$. Effectively, the torsion tensor is given by
\begin{equation}
\label{deftor}
T^{\alpha}_{\hphantom{\alpha}\mu\nu}\equiv e^{\alpha}_A \left(\partial_{\mu} e^A_{\nu} - \partial_{\nu} e^A_{\mu}\right)
\end{equation}
in terms of an orthonormal co-tetrad $e^A_{\mu}$,
$$g_{\mu\nu}=\eta_{AB}e^A_{\mu} e^B_{\nu},$$
and its matrix inverse, the tetrad $e^{\mu}_A$. In other words, both the metric $g_{\mu\nu}$ and the teleparallel connection are defined in terms of the tetrad field.

For the dynamical features, one commonly defines the superpotential or torsion-conjugate tensor as
\begin{equation}
\label{defconj}
S_{\alpha\mu\nu}\equiv\frac12 \left(T_{\alpha\mu\nu}+T_{\mu\alpha\nu}-T_{\nu\alpha\mu}\right) + g_{\alpha\mu}T_{\nu} - g_{\alpha\nu} T_{\mu}
\end{equation}
with the torsion vector 
\begin{equation}
\label{deftvec}
T_{\mu}\equiv T^{\alpha}_{\hphantom{\alpha}\mu\alpha}.
\end{equation}
In one temporal and $d$ spatial dimensions, the action functional of $f(\mathbb T)$ gravities reads
\begin{equation}
\label{defact}
S=\frac12 \int d^{d+1}x \sqrt{-g} f(\mathbb T)
\end{equation}
for the torsion scalar of the Teleparallel Equivalent of General Relativity (TEGR),
\begin{equation}
\label{deftscal}
{\mathbb T}\equiv \frac12 S_{\alpha\mu\nu}T^{\alpha\mu\nu},
\end{equation}
with the TEGR itself corresponding to the case of $f(\mathbb T)=\mathbb T$.

In general, as long as we deny any hypermomentum, the equations of motion can be presented as
\begin{equation}
\label{eom}
f^{\prime}\cdot {\mathcal G}_{\mu\nu} + \frac12 (f - f^{\prime} {\mathbb T})\cdot  g_{\mu\nu} + f^{\prime\prime}\cdot S_{\mu\nu}^{\hphantom{\mu\nu}\alpha}\partial_{\alpha}{\mathbb T} = 8\pi G \cdot \Theta_{\mu\nu}
\end{equation}
where ${\mathcal G}_{\mu\nu}$ is the standard Einstein tensor for the metric $g_{\mu\nu}$ and $ \Theta_{\mu\nu}= \Theta_{\nu\mu}$ is the energy-momentum tensor of matter fields, if any. The equation is quite complicated and prone to many well-posedness issues, see Ref. \cite{meiss} and references therein as well as our Section 7.3. In this paper, we will consider only the simplest anisotropic cosmologies of the Bianchi I type.

Our presentation will be structured as follows. In the next Section 2, we exhibit the main geometric quantities and combine them into the generalised Friedmann equations for cosmology. Vacuum solutions and cosmologies with a perfect fluid are studied in Sections 3 and 4 respectively. Then we review other works on such cosmologies in Section 5 and present the basics of Hamiltonian analysis in Section 6. After the positive message of good analytical tractability of these models, in Section 7 we turn to showing that the freedom of choosing non-diagonal tetrads makes the teleparallel structure pretty much unpredictable, adding to a long list of foundational issues in modified teleparallel gravity. Finally, in Section 8 we conclude.

\section{The cosmological model}

The spacetime is supposed to be of $1+d$ signature, with an arbitrary natural $d>1$. For the sake of convenience, it will be treated in the "mostly plus" form, with the diagonal Bianchi I configuration
\begin{equation}
\label{metrI}
g_{\mu\nu}dx^{\mu} dx^{\nu} = -N^2(t)\cdot dt^2 + \sum_i a^2_i (t)\cdot dx_i^2
\end{equation}
being represented by a diagonal co-tetrad
\begin{equation}
\label{cosm}
e^A_{\mu}={\mathrm{diag}}(N(t), a_i(t)).
\end{equation}
Since we do not use different symbols for the tetrad and co-tetrad, we will denote the different components as follows. The coordinate indices $\mu,\nu,\alpha$ will take the value $0$ for the temporal index with $i,j,k$ for the spatial ones, while for the tangent space indices $A,B$ it will be $\emptyset$ and $a,b$ respectively. So that now we have
$$e^{\emptyset}_0=N(t),\qquad e^a_0=e^{\emptyset}_i=0,\qquad e^a_i (t) = a_i(t)\cdot \delta^a_i .$$

For the tetrad (\ref{cosm}) and the metric (\ref{metrI}), the calculations are quite simple in arbitrary dimensions. As to the matter content, in the general equations, we will put the most general energy-momentum tensor with different pressures $p_i$ along the directions of $x^i$. In this case, without knowing the properties of matter, the equations do not tell us practically anything about the spacetime dynamics. For more meaningful messages, we will use either vacuum or an isotropic perfect fluid.

\subsection{Torsion quantities}

We immediately find the following torsion quantities: the torsion tensor (\ref{deftor})
$$T_{0\mu\nu}=T_{\mu 00}=T_{ijk}=0, \quad T_{i0j}=-T_{ij0}=a_i{\dot a}_i\cdot \delta_{ij};$$
the torsion vector (\ref{deftvec})
$$T_i=0, \quad T_0=\sum_i \frac{{\dot a}_i}{a_i};$$ 
the torsion-conjugate tensor (\ref{defconj})
\begin{equation}
\label{torsconj}
S_{0\mu\nu}=S_{\mu 00}=S_{ijk}=0, \quad S_{ij0}=-S_{i0j}=\delta_{ij}\cdot  a_i^2 \sum_{k\neq i} \frac{{\dot a}_k}{a_k};
\end{equation}
and finally the torsion scalar (\ref{deftscal})
\begin{equation}
\label{torsscal}
{\mathbb T}=T_{i0j}S^{i0j}=\frac{1}{N^2}\sum_i \frac{{\dot a}_i}{a_i} \sum_{j\neq i} \frac{{\dot a}_j}{a_j}.
\end{equation}

Obviously, one can also write the torsion scalar (\ref{torsscal}) as
$${\mathbb T}=\frac{1}{N^2}\left(\left(\sum_i \frac{{\dot a}_i}{a_i}\right)^2- \sum_i \frac{{\dot a}^2_i}{a^2_i}\right) =\frac{1}{N^2}\left( \left({\mathrm{Tr}}\left\{\frac{\dot a}{a}\right\} \right)^2 - {\mathrm{Tr}}\left\{\frac{\dot a}{a}\right\}^2 \right)$$
or, in a more algebraic way,
$${\mathbb T}= \frac{1}{N^2}\sum_{\substack{i,j\\ i\neq j}}\frac{{\dot a}_i{\dot a}_j}{a_i a_j} =\frac{2}{N^2}\sum_{\substack{i,j\\ i<j}}\frac{{\dot a}_i{\dot a}_j}{a_i a_j}=\frac{2}{N^2}\cdot {\mathfrak e}_2\left(\left\{\frac{\dot a}{a}\right\}\right) =\frac{{\mathfrak e}_2\left({\mathfrak g}^{-1}{\dot{\mathfrak g}}\right)}{2N^2},$$
in terms of the second elementary symmetric polynomial ${\mathfrak e}_2$ of eigenvalues of the matrix $g^{ik}{\dot g}_{kj}$. In the end of this paper we will see that the relations like
\begin{equation}
\label{badrel}
2N^2 {\mathbb T}={\mathfrak e}_2\left({\mathfrak g}^{-1}{\dot{\mathfrak g}}\right),
\end{equation}
of the teleparallel quantities to the spatial metric $\mathfrak g$, are much more general than anything what could be called a nice property of a  modified teleparallel gravity in which the teleparallel connection is supposed to be something physical\footnote{A comment is in order. One might often hear that a teleparallel connection is pure gauge. This is not true unless we are in TEGR. What is correct is that one can formally treat the spin connection components as pure gauge since they can be transformed to zero, as we have immediately done. However, the tetrad then bears information on both the metric and the independent connection, while neither geometry nor physical equations in general are invariant under Lorentz rotations of the tetrad alone. For the geometrical discussion, see paper \cite{megeom}.}.

Note that, given the calculations above (\ref{torsconj}), the $S_{\mu\nu\alpha}g^{\alpha\beta}\partial_{\beta}\mathbb T$ term in the equations (\ref{eom}) is automatically symmetric, hence no problem with their antisymmetric part. Following the common jargon in the field, we have got a "good tetrad" (\ref{cosm}). This is nothing but a particular case of a general statement that, if all elements of a diagonal tetrad are functions of only one of those coordinates in which it has been presented, then such a tetrad is always "good" in this sense \cite{CairoBH}. 

For the geometric intuition, this nice outcome is due to us using coordinates of Cartesian type \cite{meCart}. For instance, when constructing spherically symmetric solutions, any attempt of doing it with a tetrad diagonal in spherical "coordinates" actually amounts to a very strange construction of a very singular teleparallel connection on the manifold, and then the theory almost restricts itself to TEGR ($f^{\prime\prime}\partial\mathbb T =0$) that does not care about the teleparallel structure at all. However, in Cartesian coordinates it works well with a simple diagonal tetrad \cite{meCart}.

\subsection{Riemannian quantities}

The Levi-Civita (LC) connection components for the metric (\ref{metrI}) are also elementary, and they produce the Riemannian geometry with the Ricci tensor components
$${\mathcal R}_{00}=\sum_i\left(-\frac{{\ddot a}_i}{a_i} + \frac{{\dot N}{\dot a}_i}{N a_i}\right), \quad {\mathcal R}_{0i}=0, \quad {\mathcal R}_{ij}= \delta_{ij}\frac{a^2_i}{N^2}\left(\frac{{\ddot a}_i}{a_i} - \frac{{\dot N}{\dot a}_i}{N a_i} + \frac{{\dot a}_i}{a_i}\sum_{k\neq i}\frac{{\dot a}_k}{a_k}\right).$$

Note that the scalar curvature can be calculated as
$${\mathcal R} = {\mathbb T} + \frac{2}{N^2}\sum_i \left(\frac{{\ddot a}_i}{a_i} - \frac{{\dot N}{\dot a}_i}{N a_i}\right).$$
It might feel as an indicator of an unpleasant sign mistake, for we know that $\mathbb T$ as a Lagrangian density must be equivalent to $-\mathcal R$, not to $+\mathcal R$. However, one can also calculate the boundary term:
$$\bigtriangledown^{\mu}_{(\mathrm{LC})}T_{\mu}=\frac{1}{N({\mathop\Pi\limits_i}a_i)}\partial_0 \left(N({\mathop\Pi\limits_i}a_i) g^{00} T_0\right) = -{\mathbb T} -\frac{1}{N^2}\sum_i\left(\frac{{\ddot a}_i}{a_i} - \frac{{\dot N}{\dot a}_i}{N a_i}\right)$$
and see that indeed
$${\mathcal R} + {\mathbb T} + 2\bigtriangledown^{\mu}_{(\mathrm{LC})}T_{\mu} =0,$$
the teleparallel curvature vanishes and the GR and TEGR Lagrangians are equivalent.

Finally, the Einstein tensor turns out to be
\begin{equation}
\label{Einstein}
{\mathcal G}_{00}=\frac12 N^2 {\mathbb T}, \quad {\mathcal G}_{0i}=0, \quad {\mathcal G}_{ij}= -\delta_{ij} a^a_i \left(\frac12 {\mathbb T} + \frac{1}{N^2}\sum_{k\neq i} \left(\frac{{\ddot a}_k}{a_k} - \frac{{\dot N}{\dot a}_k}{N a_k} - \frac{{\dot a}_i {\dot a}_k}{a_i a_k}\right)\right).
\end{equation}
Appearance of the lapse $N$ in these expressions takes a very natural form which can be easily deduced by changing the time coordinate from the $N=1$ case.

\subsection{Generalised Friedmann equations}

Combining all the results (\ref{torsconj}, \ref{torsscal}, \ref{Einstein}) in the equations (\ref{eom}), we find that the Friedmann equations are
\begin{equation}
\label{Fried1}
{\mathbb T}f^{\prime} - \frac12 f = 8\pi G  \rho,
\end{equation}
\begin{equation}
\label{Fried2}
\frac12 f -\frac{f^{\prime}}{N^2}\left(\sum_{k\neq i}\left(\frac{{\ddot a}_k}{a_k} - \frac{{\dot N}{\dot a}_k}{N a_k} \right)- \frac{{\dot a}_i}{a_i}\sum_{k\neq i} \frac{{\dot a}_k}{a_k} + N^2 {\mathbb T}\right) - \frac{1}{N^2} \left(\sum_{k\neq i} \frac{{\dot a}_k}{a_k} \right)  f^{\prime\prime} {\dot{\mathbb T}} = 8\pi G P_i
\end{equation}
where $P_i$ is the pressure in the $i$-th direction, in case of any anisotropy.

In the gauge of $N=1$, we denote the standard Hubble parameters as
$$H_i \equiv \frac{{\dot a}_i}{a_i}$$
and the equations (\ref{Fried1}, \ref{Fried2}) take the form of
$${\mathbb T} = 2(H_1 H_2 + H_1 H_3 + H_2 H_3 + \ldots ), \qquad 2{\mathbb T}f^{\prime} - f = 16\pi G  \rho$$
and
$$- \left({\dot H}_2 + H^2_2 + {\dot H}_3 + H^2_3 + \ldots - H_1 \left(\vphantom{\int}H_2 + H_3 + \ldots\right) + {\mathbb T}\right)f^{\prime} + \frac12 f -  \left(\vphantom{\int}H_2 + H_3 + \ldots \right) f^{\prime\prime} {\dot{\mathbb T}}=  8\pi G P_1$$
with analogous expressions in other spatial directions.

We would like to show that the system of equations is actually quite tractable analytically. Using the temporal equation (\ref{Fried1}), one can immediately rewrite the spatial ones (\ref{Fried2}) as
\begin{equation}
\label{Li}
-\frac{f^{\prime}}{N^2}\left(\sum_{k\neq i}\left(\frac{{\ddot a}_k}{a_k} - \frac{{\dot N}{\dot a}_k}{N a_k} \right)- \frac{{\dot a}_i}{a_i}\sum_{k\neq i} \frac{{\dot a}_k}{a_k} \right) - \frac{1}{N^2} \left(\sum_{k\neq i} \frac{{\dot a}_k}{a_k} \right)  f^{\prime\prime} {\dot{\mathbb T}} = 8\pi G (\rho + P_i),
\end{equation}
or in explicit shapes like
$$- \left({\dot H}_2 + H^2_2 + {\dot H}_3 + H^2_3 + \ldots - H_1 \left(\vphantom{\int}H_2 + H_3 + \ldots\right) \right)f^{\prime}  -  \left(\vphantom{\int}H_2 + H_3 + \ldots \right) f^{\prime\prime} {\dot{\mathbb T}}=  8\pi G (\rho + P_1)$$
for the gauge of $N=1$. From now on we fix this gauge choice.

Let us denote the left hand sides of gravity equations (\ref{Fried1}, \ref{Li}) as ${\mathfrak L}_{\mu}$ so that
$${\mathfrak L}_0 = 8\pi G \rho, \qquad {\mathfrak L}_i = 8\pi G (\rho +  P_i).$$
One can easily see that
\begin{equation}
\label{conserv}
\sum_i H_i {\mathfrak L}_i \equiv - {\dot{\mathfrak L}}_0.
\end{equation}
On one hand, the $\sum\limits_i H_i {\mathfrak L}_i $ combination of equations (\ref{Li}) can be solved for $\dot{\mathbb T}$, as long as $f^{\prime} + 2{\mathbb T} f^{\prime\prime} \neq 0$. On the other hand, due to this identity (\ref{conserv}), we can safely analyse the spatial equations (\ref{Li}) only, and then the temporal equation (\ref{Fried1}) will just be an extra constraint on the initial data. This is actually the Bianchi identity \cite{meBian} entailing the conservation law
$${\dot\rho} + \sum_i H_i (\rho + P_i) =0.$$

First of all, we need to solve the ${\mathfrak L}_i$-equations (\ref{Li}) for the highest derivatives ${\dot H}_i$. By taking combinations of ${\mathfrak L}_i - \frac{1}{d-1} \sum\limits_j {\mathfrak L}_j$, we get
\begin{equation}
\label{spateq}
\left({\dot H}_i + H_i \sum_j H_j - \frac{\mathbb T}{d-1}\right) f^{\prime} + H_i f^{\prime\prime} \dot{\mathbb T}=-\frac{8\pi G}{d-1}\left(\rho+\sum_j P_j - (d-1)P_i\right).
\end{equation}
Now, as long as we know ${\mathbb T}(t)$, and if $f^{\prime}\neq 0$, these equations uniquely determine the accelerations ${\dot H}_i$. Anisotropic matter fluid is considered for completeness only. We will use the equations (\ref{spateq}) either in vacuum or for an isotropic perfect fluid in which case the expression in the brackets on the right hand side is simply $\rho+P$.

\subsection{The action principle in the minisuperspace}

For later use in the Hamiltonian section, note that by substituting the tetrad Ansatz (\ref{cosm}) into the action functional (\ref{defact}), we get the action of
\begin{equation}
\label{mss}
S=\frac12 \int d^{d+1}x N({\mathop\Pi\limits_i}a_i) \cdot f\left(\frac{1}{N^2}\sum_i \frac{{\dot a}_i}{a_i} \sum_{j\neq i} \frac{{\dot a}_j}{a_j}\right).
\end{equation}
One can easily see that $\delta N$- and $\delta a_i$-variations of the action (\ref{mss}) reproduce the left hand sides of the Friedmann equations above (\ref{Fried1}, \ref{Fried2}). As always \cite{meK}, fixing the gauge of $N=1$ directly inside the functional leads to the same equations (\ref{Fried2}) without the constraint (\ref{Fried1}).

\section{Vacuum cosmologies}

At first, we consider the model in vacuum, ${\mathfrak L}_{\mu}=0$. From the combination of equations above (\ref{conserv}),
$$0=\sum_i H_i {\mathfrak L}_i = -\left(\frac12 f^{\prime} + {\mathbb T} f^{\prime\prime}\right){\dot{\mathbb T}},$$ 
we see that $\dot{\mathbb T}=0$ for any function with $f^{\prime} + 2{\mathbb T} f^{\prime\prime} \neq 0$. It means that the spatial equations (\ref{spateq}) have already been solved for the highest derivatives, and they need $d$ initial data in terms of $H_i$, or $2d$ of them in terms of $a_i$. The temporal equation (\ref{Fried1}), ${\mathfrak L}_0 =0$, imposes then one constraint on these initial data by requiring a particular value of $\mathbb T$.

Since ${\mathbb T}$ is constant, the solutions are the same as in GR.\footnote{As long as $f^{\prime}\neq 0$, for otherwise anything with the corresponding value of ${\mathbb T}$ goes.} In particular, the Kasner solutions, $a_i\propto t^{p_i}$, with $\sum\limits_i p_i=\sum\limits_i p^2_i=1$, are immediately obvious when $f(0)=0$. Indeed, in this case, a possible solution for ${\mathfrak L}_0=0$ is ${\mathbb T}=0$, and then the spatial equations (\ref{spateq}) take the form of ${\dot H}_i + H_i \sum\limits_j H_j =0$, as long as $f^{\prime} \neq 0$, with the constraint $\sum\limits_i H_i^2 = \left(\sum\limits_i H_i\right)^2$ from the $\mathbb T=0$ condition. For the volume variable $V\equiv\sum\limits_i H_i$, we get an equation $\dot V = -V^2$. Putting the singularity to $t=0$, it yields $V=\frac{1}{t}$ and ${\dot H}_i =-\frac{H_i}{t}$. Therefore, $H_i=\frac{p_i}{t}$, i.e. $a_i\propto t^{p_i}$, with $\sum\limits_i p_i=1$, and $\sum\limits_i p^2_i=1$ due to the constraint.

Note that here are precisely $2d-1$ Cauchy data as it must be. Indeed, modulo some obvious inequality restrictions, we have got $d$ arbitrary constants measuring the scales of all the scale factors: $a_i=c_i t^{p_i}$, then $d-2$ arbitrary numbers in choosing the powers $p_i$, and one more datum which we have used in the origin of time: the initial singularity at $t_0=0$ can be shifted to any other instant of time.

\subsection{Solutions with a cosmological constant}

In cases of constant ${\mathbb T}\neq 0$, the solutions are much less nice but can also be easily found. By analogy with GR and a cosmological constant, we denote ${\mathbb T}=2\Lambda$ and get the equation
\begin{equation}
\label{BiaSitEq}
{\dot H}_i + H_i \sum\limits_j H_j =\frac{2\Lambda}{d-1}.
\end{equation}
Its trace for the volume variable
$$V\equiv \sum_i H_i, \qquad {\dot V} + V^2 = \frac{2d}{d-1}\Lambda\equiv C$$
provides us with the following solutions 
\begin{equation}
\label{BiaSitSol}
\frac{V}{\sqrt{C}}=\frac{e^{2\sqrt{C}t}+1}{ e^{2\sqrt{C}t}-1} \qquad \mathrm{or} \qquad \frac{V}{\sqrt{-C}}= \tan (-\sqrt{-C}t)
\end{equation}
depending on the sign of $C$, and with an integration constant eaten up by the choice of the origin of time. Then the equations (\ref{BiaSitEq})
$${\dot H}_i + H_i V(t) = \frac{C}{d}$$
can easily be solved. For example, in the case of positive $C$ solution (\ref{BiaSitSol}), one can use an integrating factor to get
$$\frac{d}{dt}\left(H_i {\mathcal V}\right)=\frac{C}{d}{\mathcal V}\qquad \mathrm{for} \qquad {\mathcal V}(t)=\left(1-e^{-2 \sqrt{C}t} \right)^{\frac{1}{\sqrt{C}}} e^t.$$
The final solutions are subject to the constraints of ${\mathbb T}=\frac{d-1}{d}C=2\Lambda$ and the value of $V(t_0)$.

To get a better feeling of what is going on, let us parametrise the anisotropic universe via 
$$H_i = H\cdot (1+\epsilon_i)\qquad \mathrm{with}\qquad \sum\limits_i \epsilon_i =0.$$
In this case we find the main variables as
\begin{equation}
\label{anispar}
V=dH, \qquad \frac{d-1}{d}C={\mathbb T}=H^2 \left(d(d-1) + \sum_{\substack{i,j\\ i\neq j}} \epsilon_i \epsilon_j\right) = H^2 \left(d(d-1) - \sum_i \epsilon_i^2\right).
\end{equation}
Therefore,
$${\mathbb T}=\frac{d-1}{d} V^2 \mathrm{\ in\ case\ of\ isotropy\ and\ } {\mathbb T}<\frac{d-1}{d} V^2 \mathrm{\ with\ any\ anisotropy}$$
what means $V^2 \geqslant C$. This is why we have chosen (\ref{BiaSitSol}) only one of two possible solutions for $V$ with positive $C$ and neglected another one, $\frac{V}{\sqrt{C}}=\frac{ e^{2\sqrt{C}t}-1}{ e^{2\sqrt{C}t}+1}$.

In the case of negative $\Lambda$, the solution (\ref{BiaSitSol}) for $V$ is very fast in developing singularities. It is not surprising due to an attempt of having flat spatial sections with negative energy density. The positive $\Lambda$ solution (\ref{BiaSitSol}) has two pieces: starting from singularity at $t=0$ and tending to isotropic expansion at $t=\infty$, as well as its time-reversed version from $t=-\infty$ to $t=0$, with the infinite time asymptotics being the de Sitter spacetimes of $H=\pm\sqrt{\frac{2\Lambda}{d(d-1)}}$.

\subsection{On the $\sqrt{\mathbb T}$ and $\Lambda+\sqrt{\mathbb T}$ models}

The model of 
$$f(\mathbb T)=\sqrt{\mathbb T}$$
is special for $\sum\limits_i H_i {\mathfrak L}_i\equiv 0$ and ${\mathfrak L}_0\equiv 0$. However, the equation (\ref{spateq}) is also valid for it. One combination of ${\dot H_i}$ cannot be found. Though, once having taken one of them, say $\dot{\mathbb T}$, as a free choice, all $d-1$ others are known. It is an accidental gauge freedom. In vacuum, and only in vacuum, an arbitrary spatially-flat Friedmann spacetime solves the equations. At the same time, the anisotropies are not free.

Assuming that $\mathbb T\neq 0$, we can multiply the equation (\ref{spateq}) by $4{\mathbb T}^{3/2}$ and get 
\begin{equation}
\label{sqrtLeq}
2{\mathbb T}\left({\dot H}_i + H_i \sum_j H_j -\frac{\mathbb T}{d-1}\right) - H_i \dot{\mathbb T}=0.
\end{equation}
It is obviously solved by any isotropic background. Taking a small anisotropy as $H_i = H\cdot (1+\epsilon_i)$ we find that ${\mathbb T}\approx d(d-1) H^2 + 2(d-1)H^2 \sum\limits_j \epsilon_j$ and derive the equation
$${\dot\epsilon}_i-\frac{1}{d}\sum_j {\dot\epsilon}_j + dH \cdot \left(\epsilon_i-\frac{1}{d}\sum_j \epsilon_j\right)\approx0$$
which immediately shows us that small anisotropies behave as $\epsilon_i - \epsilon_j \propto \frac{1}{a^d}$.

Away from small anisotropy, we parametrise it again as $H_i = H\cdot (1+\epsilon_i)$ with $\sum\limits_i \epsilon_i =0$. Precisely as we have seen it above (\ref{anispar}), the torsion scalar $\mathbb T$ has got a quadratic correction only, so that linearly we get ${\dot\epsilon}_i + dH \epsilon_i \approx 0$ indeed. At the same time, the trace of equation (\ref{sqrtLeq}) takes the form of
$${\dot V} + V^2 - \frac{d}{d-1}{\mathbb T}-\frac{V{\dot{\mathbb T}}}{2\mathbb T}=0.$$
Since $V=dH$ and ${\mathbb T}=H^2 \left(d(d-1) -A\right)$ with $A\equiv \sum\limits_i \epsilon^2_i \geqslant 0$, we see
$${\dot A} = - 2d HA +\frac{2}{d-1} HA^2$$ 
that the anisotropies decay in an expanding universe in weak anisotropy regimes only.

Obviously, the ${\mathfrak L}_i$ equations (\ref{spateq}) are absolutely the same for the model of
$$f(\mathbb T)=\Lambda+\sqrt{\mathbb T}$$
as for the case of $f(\mathbb T)=\sqrt{\mathbb T}$. The difference is in the temporal equation (\ref{Fried1}):
$$\Lambda=-16\pi G \rho.$$
Therefore, with $\Lambda\neq 0$, vacuum Bianchi I cosmologies of the tetrad Ansatz (\ref{cosm}) are not possible at all. The only possible matter content is then a cosmological constant precisely cancelling the $\Lambda$-term. Finally, it is just equivalent to the $\sqrt{\mathbb T}$ model in vacuum.

\section{Cosmologies with a perfect fluid}

For more realistic cosmologies (\ref{spateq}), let us assume an ideal isotropic fluid
\begin{equation*}
\left({\dot H}_i + H_i V - \frac{\mathbb T}{d-1}\right) f^{\prime} + H_i f^{\prime\prime} \dot{\mathbb T}=-\frac{8\pi G}{d-1}\left(\rho+P\right)
\end{equation*}
with a constant equation of state parameter $w=\frac{P}{\rho}$. Together with the temporal equation (\ref{Fried1}), it yields a very simple system again:
\begin{equation}
\label{eqcosm}
\left({\dot H}_i + H_i V + \frac{w}{d-1} {\mathbb T}\right) f^{\prime} + H_i f^{\prime\prime} \dot{\mathbb T}-\frac{w+1}{2(d-1)}f=0.
\end{equation}
For its $2d$ initial data (in terms of $a_i$), there is one constraint (\ref{Fried1}) which relates a combination of $d$ integration constants for $H_i(t)$ to the value of energy density at a chosen moment of time.

A way of solving it is quite obvious. We take two traces of the equation (\ref{eqcosm}): the $\sum\limits_i (\mathrm{Eq.})_i$ trace
\begin{equation}
\label{TV1}
\left({\dot V} + V^2 + \frac{wd}{d-1} {\mathbb T}\right) f^{\prime} + V \dot{\mathbb T}  f^{\prime\prime} -\frac{(w+1)d}{2(d-1)}f=0
\end{equation}
and the $\sum\limits_{i\neq j} H_i (\mathrm{Eq.})_j$ one
\begin{equation}
\label{TV2}
\left(\frac12 {\dot{\mathbb T}} + (w+1) {\mathbb T} V\right) f^{\prime} + {\mathbb T} \dot{\mathbb T} f^{\prime\prime}-\frac{w+1}{2}Vf=0.
\end{equation}
If one wishes to study how a deviation from isotropy behaves, it is possible to take the combination of equations $\mathbb T\ \cdot$ Eq. (\ref{TV1})  $-\ V\ \cdot$ Eq. (\ref{TV2}), for it vanishes identically whenever ${\mathbb T}=\frac{d-1}{d}V^2$.

If $w+1=0$, we are back to the situation already discussed in the previous section. Otherwise, as long as $2{\mathbb T}f^{\prime}-f\neq 0$ again, we can solve the second equation (\ref{TV2}) for $V$ algebraically,
$$(w+1)V=\frac{f^{\prime}+2{\mathbb T}f^{\prime\prime}}{f-2{\mathbb T}f^{\prime}} {\dot{\mathbb T}},$$ 
and substitute it into the first one (\ref{TV1}). As a result, the only nontrivial part of solving the system reduces to a second-order ordinary differential equation for ${\mathbb T}(t)$
$$\frac{d}{dt}\left(\frac{(f^{\prime}+2{\mathbb T}f^{\prime\prime})f^{\prime}}{(w+1)(f-2{\mathbb T}f^{\prime})} {\dot{\mathbb T}}\right) + \frac{(f^{\prime}+2{\mathbb T}f^{\prime\prime})^2 f^{\prime}}{(w+1)^2 (f-2{\mathbb T}f^{\prime})^2} {\dot{\mathbb T}}^2 - \frac{d}{2(d-1)} \left(\vphantom{T^{T^T}_{T_T}}(w+1)f- 2w{\mathbb T}f^{\prime}\right) =0,$$
a quite complicated one in general. Note in passing that the points with $f^{\prime}+2{\mathbb T}f^{\prime\prime} =0$ are also singular for the dynamics, due to disappearance of $\ddot{\mathbb T}$.

Anyway, given some known functions ${\mathbb T}(t)$ and $V(t)$, all functions $H_i(t)$ satisfy a very simple first-order ordinary differential equation (\ref{eqcosm}),
\begin{equation}
\label{finH}
{\dot H}_i + \left(V +\frac{ f^{\prime\prime}}{f^{\prime}}  \dot{\mathbb T}\right) H_i  =\frac{(w+1)f - 2w  {\mathbb T} f^{\prime}}{2(d-1) f^{\prime}}
\end{equation}
assuming $f^{\prime}\neq 0$, of course. As we see, in many respects, these tasks are tractable analytically. 

And even for the numerics, it must be much nicer this way than directly attacking the system  of coupled equations (\ref{eqcosm}). There is a mild complication though: there are two constraints for the linear equations (\ref{finH}), with only $d-2$ independent Cauchy data, for the two more of them will have already been used in taking a solution for ${\mathbb T}(t)$ and $V(t)$. The constraints on the initial values of $H_i$ will be such as to reproduce the values of $\mathbb T$ and $V$ at this moment of time.

\subsection{The case of $f({\mathbb T})={\mathbb T}^n$}

Having got the general results, we now specify it to the power law functions, $f({\mathbb T})={\mathbb T}^n$. If $\mathbb T \neq 0$, as well as $n \neq 0$ and $n \neq \frac12$, the equations (\ref{eqcosm}, \ref{TV1}, \ref{TV2}) for the power-law $f(\mathbb T)$ gravity take a simple form of
$${\dot H}_i + \left( V +  (n-1) \frac{\dot{\mathbb T}}{\mathbb T} \right) H_i  + \frac{1}{d-1}\left(w-\frac{w+1}{2n}\right) {\mathbb T} =0,$$
$$\mathrm{with}\quad \frac{\dot{\mathbb T}}{\mathbb T}=-\frac{w+1}{n} V \quad \mathrm{and} \quad {\dot V}=\left(w-\frac{w+1}{n}\right) V^2 - \frac{d}{d-1} \left(w-\frac{w+1}{2n}\right) \mathbb T.$$

In particular, as a very simple example, one can assume
$$n=\frac{w+1}{2w}.$$
For instance, it includes a 3D radiation-dominated Universe with $w=\frac13$ in a ${\mathbb T}^2$ model. We immediately see ${\dot V}=-wV^2$ and therefore $V(t)=\frac{1}{wt}$ having put the singularity time at $t=0$. Then, $ \frac{\dot{\mathbb T}}{\mathbb T}=-2wV=-\frac{2}{t}$, hence ${}\mathbb T (t)=\frac{c}{t^2}$.  Finally, for the Hubble parameters, we have ${\dot H}_i +\frac{H_i}{t}=0$, or $H_i(t)=\frac{p_i}{t}$.  It is a power law solution, 
$$a(t)\propto t^{p_i} \quad \mathrm{with\ the\ constraints\ of} \quad \sum_i p_i = \frac{1}{w}\quad \mathrm{and} \quad \sum_{\substack{i,j\\ i\neq j}}p_i p_j = c$$
and the integration constant $c$ being related to the density of matter in the universe.

\subsection{On a static extra dimension in an accelerated universe}

We would like to discuss possible applications of $f(\mathbb T)$ gravity to physical problems. Admittedly, it is an overly optimistic stance, trying to solve any physical troubles with a model which in itself is prone to many foundational issues \cite{meiss}. However, before turning to even more problems, we will present a simple idea about using the modified Bianchi I equations. Namely, it will be about a well-known issue \cite{nostextr} with compactified extra dimensions: it is nearly impossible to have an accelerated expansion in the visible universe without violating the null energy condition and with a static extra dimension, so that possible changes of its size would not lead to variations in the effective fundamental constants.

Indeed, for the Bianchi I Ansatz (\ref{metrI}) one can easily check that, even with a different pressure $P_5$ in the extra dimension, one would need $w_5<-1$ in order to have it static while the rest of the universe is isotropic and accelerating. In particular, a de Sitter phase in the latter would require $w=-1$ and $w_5=-2$. Even though one might argue \cite{maystextr} as to how generic are the assumptions needed for making the no-go statements, it is clearly a very difficult task to construct such universe in a reasonable way. Fortunately, if we take a $\sqrt{\mathbb T}$ correction to the function $f({\mathbb T})$, it changes the anisotropic equations for pressure (\ref{Fried2}) without touching upon the energy density at all (\ref{Fried1}), so that it might certainly help with modifying the required equations of state. Note that we assume a spatially homogeneous distribution of matter, not even touching upon any kind of braneworld models, though such ideas are also possible to entertain \cite{brane1, brane2}.

All in all, for an easy try, we take a model of
$$f({\mathbb T})={\mathbb T} - c \sqrt{\mathbb T}.$$
We would like to have an isotropic 3D Universe, $H(t) \equiv H_x(t)=H_y(t)=H_z(t)$, with a static extra dimension, $H_5=0$. With the $N=1$ gauge, the equations (\ref{Fried1}, \ref{Fried2}) are
\begin{equation}
\label{extrdim}
3H^2= 8\pi G \rho, \qquad 
-2{\dot H} - 3H^2 = 8\pi G P, \qquad 
-3{\dot H} - 6H^2= \underbrace{8\pi G P - \sqrt{\frac38}cH}_{8\pi G P_{\mathrm{eff}}}.
\end{equation}
The $c\sqrt{\mathbb T}$ correction makes an efective addition to the pressure in the new spatial dimension only. We are interested in using the case of $c>0$ for making $w_\mathrm{eff}\equiv\frac{P_\mathrm{eff}}{\rho}<-1$ in the extra dimension.

Let us first take an intuitive look at what we have got for the extra-dimensional pressure (\ref{extrdim}). We see that $P_\mathrm{eff} = P -\frac{c}{8\pi G}\sqrt{\frac38} H = P - \frac{c}{8}\sqrt{\frac{\rho}{\pi G}}$, and therefore the effective extra-dimensional equation of state parameter is
$$w_\mathrm{eff}=w - \frac{c}{8\sqrt{\pi G \rho}} = w - \frac{c}{2\sqrt{6}H}.$$
An obvious requirement for stability in the gravitational sector, in terms of the usual gravitons not being ghosts\footnote{Even if we are to make peace with ghosts (see papers \cite{stVik, stVer} and references therein), observation-wise we would definitely like to have a positive effective gravitational constant, while going through $f^{\prime}=0$ is anyway problematic due to strong coupling \cite{mesing}. Moreover, if we allow for ghosts, it is probably not a big deal to have $w<-1$, to start with.}, would be $f^{\prime}>0$, hence $\sqrt{\mathbb T}>\frac{c}{2}$. It implies $H>\frac{c}{2\sqrt{6}}$ or $\rho>\frac{c^2}{64\pi G}$,
and therefore
$$w_\mathrm{eff}>w-1,$$
a very mild restriction that, in principle, should allow us to combine the static extra dimension with an accelerated expansion, even if the precise solution is somewhat unnatural.

So far so good. Now we need to see what are the actual solutions. It's important to realise that the potentially interesting ones cannot be of a constant $H$ option since otherwise it would be GR in which it is impossible. One could try to solve the equations above (\ref{extrdim}) for $\rho=-P=\Lambda$, however that would produce $H=\frac{c}{2\sqrt{6}}$ and it can immediately be seen that it corresponds to $f^{\prime}=0$ and $-\frac12 f = 8 \pi G \Lambda $. It is definitely a solution, but in a pathological case.

Other possible solutions can be obtained from the equations (\ref{extrdim}) as
$${\dot H} +3H^2 = \sqrt{\frac38} cH$$
and therefore
\begin{equation}
\label{2brextr}
H(t)=\frac{c}{2\sqrt{6} \left(1\pm e^{-\sqrt{\frac38}c(t-t_0)}\right)},
\end{equation}
with the constant $H$ solution corresponding to the limit of $c t_0=-\infty$. At the same time, the matter equation of state must be
$$w=-\frac{2\dot H}{3H^2}-1 = 1 - \frac{c}{\sqrt{6}H} = 1 - \frac{c}{2\sqrt{6\pi G \rho}},$$
or in other words
$$P(\rho) = \rho - \frac{c}{2\sqrt{6\pi G}}\sqrt{\rho}.$$

The two branches (\ref{2brextr}) correspond to $w<-1$ and $w>-1$ cases. The everywhere regular branch, $H<\frac{c}{2\sqrt{6}}$, is about antigravity ($f^{\prime}<0$) and it also needs the potentially problematic matter content of $w<-1$. Another branch, $H>\frac{c}{2\sqrt{6}}$, expands from a singular state at $t=t_0$ with $w=w_{\mathrm{eff}}=1$ to a de Sitter phase of $w\to -1$ and $w_{\mathrm{eff}}\to -2$ when $t\to\infty$. Basically, when the equation of state is about to reach the value of $w=-1$, its evolution gets frozen in the de Sitter phase. The exact pathological de Sitter of $f^{\prime}=0$ precisely separates the two, $w<-1$ and $w>-1$ regimes. It might be taken as yet another illustration for the troubles of trying to cross the phantom divide line \cite{oldVik}.

\section{Comments on available literature}

Probably, the study of anisotropic solutions in $f(\mathbb T)$ cosmology was started in the paper \cite{anistart}. Already there we see an incorrect attitude to the temporal equation (\ref{Fried1}) in vacuum, $f-2{\mathbb T}f^{\prime}=0$, as an equation for the function $f$. Similar misunderstandings are very common in other papers on the subject. The solution of $f=\sqrt{\mathbb T}$ does then produce a problematic model with an arbitrary isotropic cosmology in vacuum and no matter content possible in it. However, the function $f$ is not a variable. It is a function which defines a model we use, while the equation itself can in general be solved by a constant value of the torsion scalar $\mathbb T$, with another function $f$ defining the model.

One can, of course, go the way of reconstruction \cite{Odin}, i.e. look for a function $f$ that produces a type of evolution we would like to have, either from cosmological observations or due to any other ideas. Still, even in this case, it should not be taken as a functional equation for $f(\mathbb T)$, for it must be satisfied on a particular solution only. In isotropic vacuum cosmologies, the spatial equation follows from the temporal one. And it is precisely due to the cherished property of $f(\mathbb T)$ modifications, of not producing higher derivatives, that it is all then reduced to an equation that can na{\"i}vely be taken as just a functional equation for $f$, with no spacetime derivatives on top. It ends up in the strange situation of having the full freedom of choosing $a(t)$ and impossibility of adding matter.

Then there was another paper \cite{BIAndr} with Kasner solutions for $f({\mathbb T})={\mathbb T}^n$ analysed. Some comments are in order. On top of the usual $\sum\limits_i p_i=\sum\limits_i p^2_i=1$ case when $n=1$, they present several more options:

1) For $n=\frac12$, the case of arbitrary $p_1=p_2=p_3$. This is of course about the $\sqrt{\mathbb T}$ model allowing for an arbitrary isotropic time-dependence, no need of the power law $a_i \propto t^{p_i}$ behaviour.

2) Arbitrary values of  $\left(\sum\limits_i p_i\right)^2=\sum\limits_i p^2_i$ for $n>1$. In this case we simply have $\mathbb T=0$ and $f^{\prime}(0)=0$. This is a pathological case when absolutely anything with $\mathbb T=0$ is a solution.

3) Finally, a curious case of $\sum\limits_i p_i=2n-1$ and $\sum\limits_i p^2_i=(2n-1)^2$ for $0<n<1$. Of course, it's not a proper option since already $f^{\prime}$ is singular, so that even the workings of the variational principle become doubtful. The catch is that the authors manage to non-trivially combine the $f^{\prime}$ term with the $f^{\prime\prime}{\dot{\mathbb T}}$ one in the equations (\ref{eom}), despite having $\mathbb T=0$. They do not give any detail on how precisely they did it. However, one can divide the equation by the infinite number $f^{\prime}$. Then, for arbitrary $a_i\propto t^{p_i}$, one can see that $\frac{f^{\prime\prime}}{f^{\prime}}\dot{\mathbb T}=(n-1)\frac{\dot{\mathbb T}}{\mathbb T}=-2(n-1)$. If to take it for granted, despite the fact that for the given condition on the powers $p_i$ it amounts to $\frac{0}{0}=-2$, one can easily check that, formally, the equations are solved.

On top of all that, they considered \cite{BIAndr} the temporal equation (\ref{Fried1}) in vacuum, $2{\mathbb T}f^{\prime} - f =0$, as telling us that either $f=\sqrt{\mathbb T}$ or $\mathbb T =0$ with $f^{\prime}(0)=0$. This is why there are only the $\mathbb T =0$ cases in that paper, except for the $n=\frac12$ model. Of course, this equation might also admit other constant $\mathbb T$ solutions, as has been pointed out in Refs. \cite{Topor1, Topor2}.

Note also that, in the paper \cite{Topor2} for the case of (1+3)D with isotropic perfect fluid, they have derived a relation between the Hubble parameters, with an arbitrary integration constant, also used later \cite{PTret}. Actually, this is nothing but integration of a relation obvious from our equations (\ref{finH}). Namely, that the ratios $\frac{{\dot H}_i - {\dot H}_j}{H_i - H_j}$ are all the same, $\forall i\neq j$. We find it easier to integrate the equations directly. We would like to stress that it works in any dimension, and that it does not reduce the number of degrees of freedom. The constant in their relation is a Cauchy datum, and it works just because all the Hubble parameters satisfy a differential equation of absolutely the same shape, in terms of ${\mathbb T}(t)$ and $V(t)$. However, it does not mean that "the number of independent variables is less" \cite{PTret}.

There are very few exact solutions in the available literature, with the non-constant $\mathbb T$ configurations almost absent. A nice exception to that is the paper \cite{WGnc} where one such configuration with anisotropic matter ($w=-1$ and $w=0$ parameters) is constructed for a power-law function $f$, without even specifying it explicitly as a function of $\mathbb T$. As we have seen above, one can also find such solutions with isotropic pressure. However, in such a case, they have only got a de Sitter solution \cite{WGnc}.

Two comments on their de Sitter solution \cite{WGnc} are in order. First, formally they consider an anisotropic spacetime with different constant Hubble parameters in different directions. However, they claim that the pressure must be isotropic with $w=-1$. It is easy to see then that the Hubble parameters cannot be different. In the end of the day, it is just a vacuum GR solution with a cosmological constant. 

Second, they make a rather unexpected remark \cite{WGnc} about the allegedly problematic nature of this solution, namely that the energy density is staying constant in an expanding universe, hence energy conservation violation. Yes, this is a general property of the cosmological constant! The energy is not conserved as an integral quantity, of course. Still, it behaves the very same way as it always does in GR, or in any other diffeomorphism-invariant theory of gravity.

\section{Hamiltonian analysis}

The Hamiltonian analysis of $f(\mathbb T)$ gravity is not an easy task to do \cite{Ham1, Ham2, Ham3}. However, in the minisuperspace, it is just a mechanical system. For completeness, we would like to look at the models and their cosmological dynamics in vacuum from this vantage point, too. Except for a small subsection with one remark, we will always take the total Hamiltonian only, without extending it to the so-called extended Hamiltonian which changes the behaviour of  Lagrangian variables \cite{meHam, mesing, meK}.

\subsection{The brute-force Hamiltonian}

Let us first do the Hamiltonian analysis in the most straightforward, very nonlinear approach. Namely, we take the action (\ref{mss}) as it is. We immediately get the primary constraint of $\pi_N=0$ and find the other momenta:
\begin{equation}
\label{momenta}
\pi_i = \frac{{\mathop\Pi\limits_k}a_k}{N}\cdot \frac{f^{\prime}}{a_i} \sum_{j\neq i} \frac{{\dot a}_j}{a_j}.
\end{equation}
Inside the sum of $\sum\limits_i \pi_i {\dot a}_i$ one obviously gets the torsion scalar $\mathbb T$ and, for a generic function $f$, the total Hamiltonian
\begin{equation}
\label{nlHam}
H_T = \frac{N{\mathop\Pi\limits_k}a_k}{2}\left. \left(2f^{\prime} {\mathbb T} - f\right) \vphantom{{\frac12}_{\frac{\int}{\int}}} \right|_{\mathbb T = {\mathbb T}(\pi, a)}  + \lambda \pi_N
\end{equation}
comes to our sight.

In order to express $\mathbb T$ in terms of momenta (\ref{momenta}), we find
$$f^{\prime} \frac{{\dot a}_i}{a_i} = -\frac{N}{{\mathop\Pi\limits_k}a_k} \cdot \left(\pi_i a_i - \frac{1}{d-1}\sum_j \pi_j a_j\right)$$
from what it follows
\begin{equation}
\label{Tpirel}
{f^{\prime}}^2 {\mathbb T} = \frac{1}{({\mathop\Pi\limits_k}a_k)^2}\cdot \left(\frac{1}{d-1} \left(\sum_i \pi_i a_i\right)^2 - \sum_i \pi^2_i a^2_i\right),
\end{equation}
with no dependence on $N$ inside.

We see that, as long as $\left(2f^{\prime\prime} {\mathbb T} + f^{\prime}\right) f^{\prime} \neq 0$, it is solvable for $\mathbb T$ in a locally unique way, possibly with several options. A way to have it potentially healthy is to impose $f^{\prime}>0$ and $2f^{\prime\prime} {\mathbb T} + f^{\prime} > 0$ for all values of the argument. In particular, a model of $f({\mathbb T})= {\mathbb T} + e^{\mathbb T}$ does have the function ${f^{\prime}}^2 {\mathbb T}$ monotonously growing, and therefore uniquely solvable for $\mathbb T$. At the same time, for a model of  $f(\mathbb T) \propto \sqrt{\mathbb T}$, we get yet another primary constraint, for the momenta (\ref{momenta}), with the expression above (\ref{Tpirel}) taking a constant value, see below. Between them, there are problematic cases of $2f^{\prime\prime} {\mathbb T} + f^{\prime} =0$ at some values of $\mathbb T$ only, for the function ${\mathbb T}(\pi, a)$ gets infinite derivatives there.

At the next step, we find the secondary constraint, already familiar from Lagrangian equations (\ref{Fried1}),
$$0=\Phi\equiv \left\{\pi_N, H_T \right\}= \frac{{\mathop\Pi\limits_k}a_k}{2}\left. \left(2f^{\prime} {\mathbb T} - f\right) \vphantom{{\frac12}_{\frac{\int}{\int}}} \right|_{\mathbb T = {\mathbb T}(\pi, a)}.$$
In a manner quite natural for a time-reparametrisation invariant system, it basically tells us that the Hamiltonian (\ref{nlHam}) vanishes "on-shell". Its preservation in time does not generate any new constraint, and both constraints are of the first class in Dirac's terminology.

Of course, some demands of well-posedness are typical of all non-linear generalisations of gravity. For example, even neglecting the signs of kinetic energies, in $f(R)$ gravity we need that $f^{\prime}\neq 0$ and $f^{\prime\prime}\neq 0$, the former for the usual gravitons to be physical, and the latter for the scalaron to not be infinitely strongly coupled either. The need for $f^{\prime}\neq 0$ is quite clear in our case, too. The requirement of $f^{\prime\prime}\neq 0$ is not obvious at the level of background cosmology, except for the bore of reproducing GR solutions. However, at the level of inhomogeneous perturbations, the $f^{\prime\prime}=0$ cases would lead us to strong coupling issues in the new physics beyond GR. A special feature of $f(\mathbb T)$ gravity is that already at the level of background cosmology we get yet another condition of $2f^{\prime\prime} {\mathbb T} + f^{\prime} \neq  0$, in accordance with what we have seen in the Lagrangian approach, too.

\subsection{A note on an extended Hamiltonian}

As a brief remark, we would like to mention that it is easy to see what happens in this case when adding the secondary constraint to the Hamiltonian (\ref{nlHam}). It shifts the variable $N$ in the canonical Hamiltonian by an arbitrary Lagrange multiplier of the secondary constraint. Effectively, it means a new Hamiltonian consisting of two first-class constraints only,
$$H_E = \lambda_H \left. \left(2f^{\prime} {\mathbb T} - f\right) \vphantom{{\frac12}_{\frac{\int}{\int}}} \right|_{\mathbb T = {\mathbb T}(\pi, a)}  + \lambda \pi_N.$$

Therefore, the lapse becomes absolutely free and independent, while the scale factors are required to satisfy the usual equations with another arbitrary function of time in place of $N$, like if the time reparametrisation symmetry acted on the lapse function and on the spatial sector independently. It is similar to adding the secondary constraint to the Hamiltonian of electrodynamics that makes both temporal and longitudinal photons pure gauge while moving the physics of the Coulomb potential to the longitudinal momentum which loses its relation to the Lagrangian fields \cite{meK, mesing, meHam}.

\subsection{The Hamiltonian artificially linearised in $\mathbb T$}

In order to have a simpler Hamiltonian, for the aims of quantisation or whatever else, and assuming $f^{\prime\prime}\neq 0$, one could take the Lagrangian
$$S=\frac12 \int d^{d+1}x N({\mathop\Pi\limits_i}a_i) \cdot \left(f(T) - f^{\prime}(T)\cdot \left(T - \frac{1}{N^2}\sum_i \frac{{\dot a}_i}{a_i} \sum_{j\neq i} \frac{{\dot a}_j}{a_j}\right)\right),$$
equivalent to the initial one (\ref{mss}). We have got the same momenta as above (\ref{momenta}), with the only difference in that the argument of $f^{\prime}$ becomes an independent variable $T$, and a new primary constraint $\pi_T=0$.

The total Hamiltonian is
\begin{equation}
\label{anHam}
H_T = \frac{N{\mathop\Pi\limits_k}a_k}{2}\left(f^{\prime} {\mathbb T} + f^{\prime}  T - f\right) \vphantom{{\frac12}_{\frac{\int}{\int}}}  + \lambda_N \pi_N + \lambda_T \pi_T
\end{equation}
with the $\mathbb T$ quantity defined by
$${\mathbb T}=  \frac{1}{{f^{\prime}}^2({\mathop\Pi\limits_k}a_k)^2}\cdot \left(\frac{1}{d-1} \left(\sum_i \pi_i a_i\right)^2 - \sum_i \pi^2_i a^2_i\right).$$
Omitting the obvious prefactors, Poisson brackets of the primary constraints $\pi_N$ and $\pi_T$ with the Hamiltonian (\ref{anHam}) lead to the Hamiltonian constraint again, 
$$f^{\prime} {\mathbb T} + f^{\prime}  T - f =0,$$
and, relying on $f^{\prime\prime}\neq 0$, to another constraint\footnote{In order to arrive at this result, one needs to be careful and not forget that the variable $T$ also enters the quantity $\mathbb T$ through the function ${f^{\prime}}^2$.} of 
$${\mathbb T}=T$$
respectively. The latter secondary constraint is of second class with $\pi_T$, while the former one is of the first class together with $\pi_N$. The constraint analysis stops here. No new constraints are produced, for the Hamiltonian constraint is automatically preserved, while preservation of the $T=\mathbb T$ constraint fixes the value of the Lagrange multiplier $\lambda_T$ as is always the case with second class constraints.

\subsection{On the $\sqrt{\mathbb T}$ and $\Lambda+\sqrt{\mathbb T}$ models again}

The special feature of the 
$$f({\mathbb T})=\sqrt{\mathbb T}$$ 
model (necessarily in vacuum) is in the absence of the Hamiltonian constraint. Indeed, either the canonical Hamiltonian is identically zero in the brute-force approach (\ref{nlHam}) or its vanishing immediately follows from the $T=\mathbb T$ constraint with no further restriction coming from it (\ref{anHam}). An interesting consequence of this fact is that the lapse gets totally decoupled: ${\dot \pi}_N=0$ and ${\dot N}=\lambda_N$, with no influence on the equations for scale factors $a_i$.

One can easily understand the reason. For the metric (\ref{metrI}) and the torsion scalar (\ref{torsscal}), the Lagrangian density $\sqrt{-g}\sqrt{\mathbb T}$ does not depend on the lapse at all. It is a trivial case of gauge symmetry when we put a variable to the arguments of the Lagrangian function only formally, while in reality it is simply not there, and does not produce any further constraints on the remaining sector of the model \cite{meK}. Since it was a part of a nontrivial gauge symmetry, the latter has to be there in the form of a new first class primary constraint (\ref{Tpirel}),
$$\frac{1}{d-1} \left(\sum_i \pi_i a_i\right)^2 - \sum_i \pi^2_i a^2_i - \frac14 \left({\mathop\Pi\limits_k}a_k\right)^2=0.$$
Even though of no Lagrangian constraint, it corresponds to a gauge freedom of isotropic spacetimes. Substituting a spatially-flat Friedmann universe of the tetrad (\ref{cosm}) into the action (\ref{defact}, \ref{mss}) for $f({\mathbb T})=\sqrt{\mathbb T}$, one can see that the Lagrangian density is a full time derivative \cite{BIAndr}.

We can now remove the trivial lapse variable from the model completely and have the total Hamiltonian of
$$H_T = \lambda \left(\frac{1}{d-1} \left(\sum_i \pi_i a_i\right)^2 - \sum_i \pi^2_i a^2_i - \frac14 \left({\mathop\Pi\limits_k}a_k\right)^2 \right).$$
Since the primary constraint is the only non-zero part of the whole thing, the constraint analysis obviously stops here. For the illustration purposes, let us see how the Bianchi I dynamical equations can be reproduced in the Hamiltonian approach.

The Hamiltonian equations are
$${\dot a}_i=2\lambda \left(\frac{a_i}{d-1}  \sum_j \pi_j a_j -  \pi_i a^2_i \right), \qquad {\dot\pi}_i =-2\lambda \left(\frac{\pi_i}{d-1} \sum_j \pi_j a_j -  \pi^2_i a_i - \frac{1}{4a_i} \left({\mathop\Pi\limits_k}a_k\right)^2 \right),$$
and we deduce $\frac{d}{dt}(\pi_i a_i)=\frac{\lambda}{2} \left({\mathop\Pi\limits_k}a_k\right)^2 $, with no summation over $i$. If we define a variable
$$H_i\equiv\frac{{\dot a}_i}{a_i}= 2\lambda \left(\frac{1}{d-1}  \sum_j \pi_j a_j -  \pi_i a_i \right),$$
then, by direct differentiation, we find the dynamical equation for it
$${\dot H}_i=\frac{\lambda^2}{d-1}  \left({\mathop\Pi\limits_k}a_k\right)^2 +\frac{\dot\lambda}{\lambda} H_i.$$
Upon introducing a new arbitrary function ${\mathbb T}(t) \equiv \lambda^2  \left({\mathop\Pi\limits_k}a_k\right)^2$, we get $2\frac{\dot\lambda}{\lambda} + 2V = \frac{\dot{\mathbb T}}{\mathbb T}$ with $V\equiv\sum\limits_k H_k$, and the Lagrangian equations (\ref{sqrtLeq})
$${\dot H}_i = \frac{\mathbb T}{d-1} + H_i\left( \frac{\dot{\mathbb T}}{2\mathbb T} - V\right)$$
are reproduced indeed.

Finally, the model of
$$f({\mathbb T})=\Lambda + \sqrt{\mathbb T}$$ 
in vacuum gets the Hamiltonian
$$H_T = -\frac{\Lambda N}{2} {\mathop\Pi\limits_k}a_k +\lambda_N \pi_N + \lambda \left(\frac{1}{d-1} \left(\sum_i \pi_i a_i\right)^2 - \sum_i \pi^2_i a^2_i - \frac14 \left({\mathop\Pi\limits_k}a_k\right)^2 \right).$$
With $\Lambda\neq 0$, the preservation of the $\pi_N=0$ constraint leads to ${\mathop\Pi\limits_k}a_k=0$, an unacceptable secondary constraint which shows that there are no solutions, as we have also seen in the Lagrangian approach.

\section{Non-diagonal tetrads for Bianchi I spacetimes}

The most general Bianchi I configuration in physical time is
\begin{equation}
\label{genBia}
g_{\mu\nu}dx^{\mu} dx^{\nu} = - dt^2 + g_{ij}(t) dx^i dx^j.
\end{equation}
Without using dynamical equations of GR, or probably of some other model, it cannot be diagonalised. Indeed, without specifying dynamics, the matrix $g_{ij}$ is a symmetric matrix of arbitrary dependence on time. At every instant of time, the matrix is diagonalisable for sure, but the eigenvectors might also depend on time, so that any attempt of diagonalising it everywhere would lead to non-vanishing mixed terms, $dt\, dx^i$.

It is of interest for us now since, given a non-diagonal Bianchi I metric, there is no natural choice of the tetrad because a diagonal one is not available. Therefore, one is led to investigating more general tetrad choices. In nontrivial modifications of TEGR, it is something worth being analysed. To start with, we review the case of GR. Unfortunately, it is rarely explained in the literature why one can always choose coordinates in which a Bianchi I universe in GR enjoys a diagonal metric.

\subsection{Brief remarks on Bianchi I in GR}

 Let us denote the spatial metric (\ref{genBia}) as a matrix $\mathfrak g$. Another notation will be a matrix 
$${\mathfrak M}\equiv {\mathfrak g}^{-1}{\dot{\mathfrak g}}$$
and the $\left[\mathfrak M\right]$ symbol representing the trace of $\mathfrak M$. In other words, ${\mathfrak M}^i_j=g^{ik}{\dot g}_{kj}$ and $\left[\mathfrak M\right]=g^{ik}{\dot g}_{ki}$. Then the Ricci tensor takes the form of
$${\mathcal R}_{00}=-\frac12  \dot{\left[{\mathfrak M}\right]} -\frac14 \left[{\mathfrak M}^2\right]=-\frac12 \left[{\mathfrak g}^{-1}{\ddot{\mathfrak g}}\right] + \frac14 \left[{\mathfrak M}^2\right] , \qquad {\mathcal R}_{0i}=0, \qquad {\mathcal R}_{ij}= \frac12 \ddot{\mathfrak g} +\frac14 \left[{\mathfrak M}\right] \dot{\mathfrak g} - \frac12 \dot{\mathfrak g} {\mathfrak M}$$
where the elementary fact of ${\dot{\mathfrak M}}={\mathfrak g}^{-1} \ddot{\mathfrak g} - {\mathfrak M}^2$, based on $\frac{d}{dt} {\mathfrak g}^{-1} = - {\mathfrak g}^{-1} {\dot{\mathfrak g}} {\mathfrak g}^{-1}$,  has been used.
Note that the marix $\mathfrak M$ does not need to be symmetric, even though the matrix $\dot{\mathfrak g} {\mathfrak M}$ is automatically symmetric:
$$\left(\dot{\mathfrak g} {\mathfrak M}\right)^T = \left(\dot{\mathfrak g} {\mathfrak g}^{-1}{\dot{\mathfrak g}} \right)^T = {\dot{\mathfrak g}}^T \left({\mathfrak g}^{-1}\right)^T{\dot{\mathfrak g}}^T = \dot{\mathfrak g} {\mathfrak g}^{-1}{\dot{\mathfrak g}}  = \dot{\mathfrak g} {\mathfrak M}$$
where the upper index ${}^T$ is the transposition sign.

For a solution in vacuum, we need
\begin{equation}
\label{1eq}
{\mathcal G}_{00} = \frac18\left(\left[{\mathfrak M}\right]^2 - \left[{\mathfrak M}^2\right] \right) = 0
\end{equation}
and
\begin{equation}
\label{2eq}
{\mathcal R}^i_j = \frac12 \left({\dot{\mathfrak M}}^i_j + \frac12  \left[{\mathfrak M}\right] {\mathfrak M}^i_j \right) =0.
\end{equation}
The first equation (\ref{1eq}) tells us that the second elementary symmetric polynomial of eigenvalues of the matrix $\mathfrak M$ must vanish, while the trace of the second (\ref{2eq}) equation ${\dot{\mathfrak M}} = - \frac12  \left[{\mathfrak M}\right] {\mathfrak M}$  demands that $\left[{\mathfrak M}\right]=\frac{2}{t}$ where the time of singularity is chosen to be at $t=0$. The resulting equation ${\dot{\mathfrak M}} = - \frac{\mathfrak M}{t}$ is easy to solve:
$${\mathfrak M}(t)=\frac{{\mathfrak M}_0}{t}, \qquad \left[{\mathfrak M}_0\right]=2, \qquad \left[{\mathfrak M}^2_0\right]=4$$
with a constant matrix ${\mathfrak M}_0$.

Note that if we assume the matrix ${\mathfrak M}_0$ to be symmetric, then it means that the matrix $\mathfrak M$ will stay symmetric always. In this case ${\dot{\mathfrak g}} {\mathfrak g}^{-1} = {\dot{\mathfrak g}}^T {{\mathfrak g}^{-1}}^T = {\mathfrak M}^T = {\mathfrak M} = {\mathfrak g}^{-1}{\dot{\mathfrak g}}$. In other words, the matrix $\mathfrak M$ is symmetric if and only if the matrices $\mathfrak g$ and $\dot{\mathfrak g}$ commute with each other, and therefore they all commute and are simultaneously diagonalisable. The equation for ${\mathfrak g}(t)$ can then  easily be solved as ${\mathfrak g} = {\mathfrak g}_0 t^{{\mathfrak M}_0}$ in the standard meaning of functions of self-adjoint operators. And we are back to the traditional diagonal Bianchi I spacetimes. Indeed, having chosen the values of ${\mathfrak g}_0$ and $\dot{\mathfrak g}_0$ diagonal at the time $t=1$, we get them diagonal always.

At the same time, there are definitely solutions with the matrix ${\mathfrak M}_0$ non-symmetric. Actually, they can be brought to the diagonal form by a non-orthonormal change of spatial variables. Indeed, assume that we have diagonalised the matrix ${\mathfrak g}_0$. Then, having got ${\mathfrak g}_0 =\mathrm{diag}(g_{(1)}, g_{(2)}, \ldots, g_{(d)})$, we can change the variables to $X^i = \sqrt{g_{(i)}}\cdot x^i$ and reduce the initial datum to ${\mathfrak g}_0 = I$, the unit matrix. Now, by an orthonormal transformation of the new variables, it's possible to diagonalise the initial velocity ${\dot{\mathfrak g}}_0$, too\footnote{This is somewhat reminiscent of the possibility to trivialise a given metric at a chosen point $x_*$, up to the first order around the point: $g_{\mu\nu}(x)=\eta_{\mu\nu} + {\mathcal O}\left((x-x_*)^2\right)$.}. As the Einstein equations are of second order at most, this is enough for being able to diagonalise the Bianchi I spacetimes in GR.

Since the full coordinate change to the diagonal representation is not orthonormal, one might be interested in studying the teleparallel configurations with non-diagonal Bianchi I. However, for the sake of clarity, let us first describe a simple example.

\subsection{An example}

Consider the following initial data at the time $t_0=1$:
$$ {\mathfrak g}_0=
\begin{pmatrix}
6 & 0& 0\\
0 & 2 & 0\\
0 & 0 & 2
\end{pmatrix},
\qquad {\dot{\mathfrak g}}_0=
\begin{pmatrix}
3 & 3& 0\\
3 & 0 & 0\\
0 & 0 & 3
\end{pmatrix}.
$$
Those are not simultaneously diagonalisable. The matrix
$${\mathfrak M}_0 =
\begin{pmatrix}
\frac16 & 0& 0\\
0 & \frac12 & 0\\
0 & 0 & \frac12
\end{pmatrix}
\times
\begin{pmatrix}
3 & 3& 0\\
3 & 0 & 0\\
0 & 0 & 3
\end{pmatrix}
=
\begin{pmatrix}
\frac12 & \frac12& 0\\
\frac32 & 0 & 0\vphantom{\frac{\int}{\int}}\\
0 & 0 & \frac32
\end{pmatrix}
 $$
has got $\left[{\mathfrak M}_0\right]=2$ and $\left[{\mathfrak M}^2_0\right]=4$, so that the necessary constraints are satisfied.

All we need is to solve an equation
$$\dot{\mathfrak g} = \mathfrak{gM} = \frac{\mathfrak g}{t} \times 
\begin{pmatrix}
\frac12 & \frac12& 0\\
\frac32 & 0 & 0\vphantom{\frac{\int}{\int}}\\
0 & 0 & \frac32
\end{pmatrix}$$
with the initial (at $t=1$) conditions from above. It is easy to do so in components and get
$$g_{13}=0, \qquad g_{13}=0, \qquad g_{33} = 2 t^{3/2},$$
$$g_{22}=C_1 t^{\alpha_1} + C_2 t^{\alpha_2}, \qquad g_{12} = 2C_1 \alpha_1 t^{\alpha_1} + 2C_2 \alpha_2 t^{\alpha_2}, \qquad g_{11}=3g_{22}+g_{12}$$
with 
$$\alpha_{1,2}=\frac14 \left(1\pm \sqrt{13}\right), \qquad C_1 + C_2 = 2, \qquad C_1 \alpha_1 + C_2 \alpha_2 = 0.$$

Even though the solution is obviously not diagonalisable by an orthonormal rotation of coordinate axes, one can see that the metric takes the form of
$$C_1 t^{\alpha_1}\left(dy + 2\alpha_1 dx\right)^2 + C_2 t^{\alpha_2}\left(dy + 2\alpha_2 dx\right)^2 + 2t^{3/2} dz^2,$$
and is therefore equivalent to a diagonal one in somewhat different coordinates.

\subsection{A problem with $f(\mathbb T)$ gravity}

What we can see now is that an arbitrary tetrad $e^A_{\mu}$ of the form 
\begin{equation}
\label{newtet}
e^{\emptyset}_0=N(t),\qquad e^a_0=e^{\emptyset}_i=0,\qquad \mathrm{and\ an\ arbitrary}\  e^a_i (t) \ \mathrm{matrix}
\end{equation}
automatically solves the antisymmetric part of equations (\ref{eom}). Indeed, the torsion tensor is still reduced to the components $T^i_{\hphantom{i}0j}=-T^i_{\hphantom{i}j0} = e^i_a {\dot e}^a_j$ only, while the relevant torsion-conjugate components
$$S_{ij0}=\frac12 \left(T_{ij0}+T_{ji0}\right)+g_{ij}T_0$$
are purely symmetric, with only these and $S_{i0j}=-S_{ij0}$ components being non-zero. Hence, the crucial $f^{\prime\prime}S_{\mu\nu\alpha}\partial^{\alpha}\mathbb T$ term in the equations (\ref{eom}) is also symmetric.  

It is a serious generalisation of the theorem from the paper \cite{CairoBH} about a diagonal tetrad.  Of course, the distinguished coordinate does not need to be a time-like one for this kind of proposition to be true. Therefore, we can formulate it as yet another

{\bf Theorem.} In a given coordinate system, let there be only one coordinate $x^{\xi}$ on which the orthonormal tetrad components may depend. Let's use the letter $C$ for the tangent space index whose number corresponds to this coordinate, $C=\xi$. Then an arbitrary tetrad of the form
$$e^C_{\xi} = N(x^{\xi}), \qquad e^A_{\xi}=e^C_{\mu}=0 \ \mathrm{for}\ A\neq C \ \mathrm{and} \ \mu\neq\xi, \qquad \mathrm{and \ an \ arbitrary}\ e^A_{\mu}(x^{\xi})\ \mathrm{matrix}$$
does automatically satisfy the antisymmetric part of equations (\ref{eom}).

An important difference from the diagonal tetrad theorem of Ref. \cite{CairoBH} is that now the tensor components $S_{\mu\nu\xi}$ are only symmetric in their free indices, while there they were diagonal. One might correctly suspect that this absence of antisymmetric equations must be related to a kind of invariance under $x^{\xi}$-dependent rotations in the orthogonal hypersurface.

With the time variable in the role of the $x^{\xi}$ coordinate, the new proposition immediately shows us a severe dynamical problem with $f(\mathbb T)$ gravity. Namely, an elementary calculation for the tetrad (\ref{newtet}) in the $N=1$ gauge yields:
$$S_{ij0}=\frac12 \left( g_{ij}g^{kl}{\dot g}_{kl} -{\dot g}_{ij}\right) =\frac12 \left(\left[\mathfrak M\right]\mathfrak g - {\dot{\mathfrak g}}\right)_{ij}$$
with $g_{ij}\equiv\delta_{ab}e^a_i e^b_j$, the same as it was with the diagonal tetrad (\ref{cosm}, \ref{torsconj}), and
$${\mathbb T}= \frac14 \left(\left(g^{ij}{\dot g}_{ij}\right)^2 - {\dot g}_{ij}g^{ik}g^{jl}{\dot g}_{kl}\right)=\frac14\left(\left[{\mathfrak M}\right]^2 - \left[{\mathfrak M}^2\right] \right) =\frac12 {\mathfrak e}_2\left({\mathfrak g}^{-1}{\dot{\mathfrak g}}\right)$$
with the very same relation (\ref{badrel}) as before. In other words, the equations of motion (\ref{eom}) with an arbitrary co-tetrad (\ref{newtet}) can be written down in terms of the metric only. Therefore, the choice of such tetrad (\ref{newtet}) is absolutely free, with arbitrary time-dependent spatial rotations of the tetrad not influencing anything in the equations. In this relatively symmetric situation, the teleparallel geometry with its torsion tensor turns out to be pretty much unpredictable, and it concerns the isotropic Friedmann universes, too.

This outcome would be perfect for TEGR. It has got even much more, the full local Lorentz freedom as its gauge symmetry. Only the Riemannian geometry is the physical one in TEGR, in the sense of being predictable, with the choice of teleparallel connection being pure gauge. However, modified models are different, and what we see here is from the chaotic and problematic realms of strong couplings \cite{meiss} and remnant symmetries \cite{remnant}. Historically, the notion of remnant symmetries was formulated \cite{remnant} as those local Lorentz transformations of the tetrad which leave the torsion scalar $\mathbb T$ invariant. Both Friedmann and Bianchi I spacetimes were considered \cite{remnant}, with wide classes of such transformations. What we have seen in this section is that, when those rotations are restricted to dependence on time only, the outcome is not just a formal invariance of a scalar, instead it is a full ignorance in all the field equations of such transforms.

All the solutions from previous sections are reproduced, and only them metric-wise, however the tetrad can be changed by an arbitrary time-dependent orthonormal rotation of its spatial part. Effectively, in the Lorentz-covariant language, such solutions were mentioned\footnote{One can also find another construction there \cite{covTBian}, with a Lorentz boost along a linearly expanding direction $x$ with the $yz$ plane being isotropic. This is another interesting aspect. However, for making our point now the configurations (\ref{newtet}) are more important.} in Ref. \cite{covTBian}, without discussing the physical consequences. A really bad point about this result is that the teleparallel connection appears to be unpredictable, similar to what was observed in earlier works \cite{prob1, prob2} for a boost in a fixed direction. If it was about a stably existent gauge symmetry, it would be all right, but it is not the case \cite{prob3}. 

\section{Conclusions}

We have seen that the dynamical equations of Bianchi I spacetimes do have a very nice and analytically tractable structure, not only in GR but also in $f({\mathbb T})$ gravity models. To the best of our knowledge, for $f(\mathbb T)$ with matter, up to now there were either very special solutions obtained by a skillful construction or some numerical investigations. What we have seen in this paper is that the dynamical equations can tell us a lot, even when being analysed analytically. Actually, our approach must also be extendable to many other modified gravity theories. 

At the same time, a glance at the general nondiagonal Bianchi type I metrics has brought us to understanding that the freedom of a proper tetrad choice is too wide for these models, representing severe dynamical issues in modified teleparallel frameworks. Even in the case of isotropic cosmology, one does not need to go for the nice diagonal tetrad only. It can be changed by an arbitrary time-dependent spatial rotation. It brightly spotlights the unpredictability problem in modified teleparallel gravity once again, considerably enhancing the previous findings \cite{prob1, prob2, prob3}.

This unpredictability can also be illustrated by cosmological perturbations in $f(\mathbb T)$ gravity \cite{mecosm, mecosm2}. The spatial rotations of the tetrad enter the perturbative equations with spatial derivatives only, and their (pseudo)scalar part does not make it to the linear equations at all \cite{mecosm, mecosm2}. For sure, our unpredictability claim applies to spatially homogeneous rotations. Rotations that depend on spatial coordinates are restricted by equations, even though for the scalar part of them one would need to go for nonlinear corrections. Still, it does not make the general $f(\mathbb T)$ equations well-posed, nor the cosmological approximations in them reliable.





\end{document}